\documentclass{ws-ijmpd}

\begin{document}

\markboth{U. Mukhopadhyay, P. P. Ghosh and S. Ray} {Higher
Dimensional Dark Energy Investigation with Variable $\Lambda$ and
$G$}

%
\catchline{}{}{}{}{}
%

\title{Higher Dimensional Dark Energy Investigation with Variable $\Lambda$ and $G$}

\author{UTPAL MUKHOPADHYAY}

\address{Satyabharati Vidyapith, North 24
Parganas, Kolkata 700 126, West Bengal, India
\\e-mail: utpal1739@gmail.com}

\author{PARTHA PRATIM GHOSH}

\address{Department of Physics, A. J. C. Bose Polytechnic,                                                                             Date: 26.11.09
Berachampa, North 24 Parganas, West Bengal, India
\\e-mail: parthapapai@gmail.com}

\author{SAIBAL RAY}

\address{Department of Physics, Government College of Engineering and Ceramic Technology, Kolkata
700 010, West Bengal, India\footnote{Corresponding address} \\
e-mail: saibal@iucaa.ernet.in}

\maketitle

\begin{history}
\received{Day Month Year} \revised{Day Month Year}
\end{history}

\begin{abstract}
Time variable $\Lambda$ and $G$ are studied here under a
phenomenological model of $\Lambda$ through an ($n+2$) dimensional
analysis. The relation of Zeldovich~\cite{Zeldovich1968}
$|\Lambda| = 8\pi G^2m_p^6/h^4$ between $\Lambda$ and $G$ is
employed here, where $m_p$ is the proton mass and $h$ is Planck's
constant. In the present investigation some key issues of modern
cosmology, viz. the age problem, the amount of variation of $G$
and the nature of expansion of the Universe have been addressed.
\end{abstract}


\section{Introduction}
The inception of the idea of a possible time variation of the
gravitational constant $G$ by Dirac~\cite{Dirac1937} and
subsequent supportive works, both at
theoretical~\cite{Brans1961,Hoyle1972,Dirac1973,Marciano1984,Will1987,Goldman1992,Bertolami1993}
and
observational~\cite{Reasemberg1983,Hellings1987,Kaspi1994,Arzoumanian1995,Guenther1998,Gaztanaga2001,Turyshev2003,Stairs2003,Biesiada2004,Benvenuto2004}
level opened up a new pathway in cosmological research. Numerous
works in this direction have been done by taking $G$ as a
variable. After the discovery of the scenario of accelerating
Universe~\cite{Perlmutter1998,Riess1998}, investigations within
the framework of variable $G$ is also not uncommon in the
literature. One of the techniques of dark energy investigation is
to include a $\Lambda$ term in the field equations of Einstein.
Since a constant $\Lambda$ is handicapped by the well known
Cosmological Constant Problem and Coincidence Problem, so it is
not unnatural to consider $\Lambda$ as a variable quantity. When
the $\Lambda$ term is placed in the right hand side of the field
equations and is considered as a part of the energy-momentum
tensor, then instead of $T^{\mu\nu}$, the total energy-momentum
tensor $T^{\mu\nu}+(\Lambda/8\pi G)g^{\mu\nu}$ is conserved. On
the other hand, as an extension of Dirac's Large Number Hypothesis
(LNH), Zeldovich~\cite{Zeldovich1968} established a connection (to
be shown explicitly afterwards) between the cosmological parameter
$\Lambda$ and the gravitational constant $G$ by considering
$\Lambda$ as the gravitational energy of the vacuum. So,
variability of $\Lambda$ admits the variability of $G$ also
provided one considers the other quantities of the relation of
Zeldovich~\cite{Zeldovich1968} as constant.

As a part of dark energy investigation, Ray et al.~\cite{Ray2007a}
considered $\Lambda$ models in four dimensional space-time by
taking $G$ as time-dependent. So, it is not unnatural to make an
attempt for exploring new physical features by venturing into
dimensions higher than the usual $4D$. This is the motivation
behind the present work where a $(n+2)$ dimensional study of
time-dependent $\Lambda$ model has been done within the framework
of variable $G$. We have addressed here some key issues of modern
cosmology, viz. the age problem, the amount of variation of $G$
and the nature of expansion of the Universe. The scheme of the
study is as follows: Section 2 deals with field equations and
their solutions while some physical features arising out of this
investigation are described in different subsections of Section 3.
Finally, some concluding remarks are made in Section 4.

\section{Field Equations and Their Solutions}
The $(n+2)$ dimensional metric for homogeneous and isotropic
Universe is given by
\begin{eqnarray}
 ds^2 = dt^2- a^2(t)[dr^2+r^2(dx_n)^2],
\end{eqnarray}
where $a(t)$ is the scale factor and
\begin{eqnarray}
(dx_n)^2 =
d\theta^2+sin^2\theta_1d\theta_2^2+......+sin^2\theta_1^2sin^2\theta_2^2....sin^2d\theta_n^2.
\end{eqnarray}
For a flat ($k=0$) Universe, the above metric yields the field
equations given by
\begin{eqnarray}
\frac{n(n+1)}{2}\left(\frac{\dot a}{a}\right)^2 = 8\pi G\rho
+\Lambda
\end{eqnarray}
and
\begin{eqnarray}
n\frac{\ddot a}{a}+\frac{n(n-1)}{2}\left(\frac{\dot a}{a}\right)^2
= -8\pi G p+\Lambda,
\end{eqnarray}
where $\Lambda\equiv\Lambda(t)$ and $G\equiv G(t)$.

The barotropic equation of state is given by
\begin{eqnarray}
p = \omega \rho,
\end{eqnarray}
where the barotropic index $\omega$, assumed to be constant here,
can take the values $0$, $1/3$, $1$ and $-1$ for pressureless
dust, electromagnetic radiation, stiff (Zeldovich) fluid and
vacuum fluid respectively. Some other limits on $\omega$ coming
from SN Ia data~\cite{Knop2003} and a combination of SN Ia data
with CMB anisotropy and galaxy clustering
statistics~\cite{Tegmark2004} are given by $-1.67<\omega<-0.62$
and $-1.3<\omega<-0.79$ respectively.

Now, the relation of Zeldovich~\cite{Zeldovich1968} between
$\Lambda$ and $G$ is given by
\begin{eqnarray}
|\Lambda| = \frac{8\pi G^2m_p^6}{h^4},
\end{eqnarray}
where $m_p$ is the proton mass and $h$ is Planck's constant.

From equation (6) we can write
\begin{eqnarray}
G = C\sqrt\Lambda,
\end{eqnarray}
where $C = h^2/\sqrt(8\pi)m_p^3$ = constant.

Let us use the {\it ansatz}
\begin{eqnarray}
\Lambda = 3\alpha H^2,
\end{eqnarray}
where $\alpha$ is a parameter.

Then from (3) we have
\begin{eqnarray}
\rho = \left[\frac{n(n+1)-6\alpha}{16\pi B\sqrt \alpha}\right] H,
\end{eqnarray}
where $\dot a/a=H$ and $B=\sqrt3 C$.

Using equations (9) and (5) and remembering that $\ddot a/a=\dot
H+H^2$ we get from~(4),
\begin{eqnarray}
n\dot H = (1+\omega)\left[\frac{6\alpha-n(n+1)}{2}\right]H^2.
\end{eqnarray}

Solving equation (10) we get our solution set as
\begin{eqnarray}
a(t)= C_1 t^{\frac{2n}{(1+\omega)[n(n+1)-6\alpha}]},
\end{eqnarray}
\begin{eqnarray}
H(t) = \frac{2n}{(1+\omega)[n(n+1)-6\alpha]}t^{-1},
\end{eqnarray}
\begin{eqnarray}
\rho(t) = \frac{n}{8\pi B(1+\omega)\sqrt\alpha}t^{-1},
\end{eqnarray}
\begin{eqnarray}
G(t) = \frac{2nB\sqrt\alpha}{(1+\omega)[n(n+1)-6\alpha]}t^{-1},
\end{eqnarray}
\begin{eqnarray}
\Lambda(t) =
\frac{12n^2\alpha}{{(1+\omega)^2}[n(n+1)-6\alpha]^2}t^{-2},
\end{eqnarray}
where $C_1$ is a constant.

It is interesting to note that in four dimensional case (i.e.
$n=2$), expressions for $a(t)$, $H(t)$ and $\Lambda(t)$ become
identical with that of Ray et al. \cite{Ray2007a} for the same
$\Lambda$ model and under the same assumptions, i.e. by taking
both $\Lambda$ and $G$ as time-dependent but without choosing any
particular expression for $G$. Moreover, here $G\propto 1/t$ as
obtained earlier also by Dirac~\cite{Dirac1937} as well as by Ray
et al.~\cite{Ray2007a}. Here, for physical validity, $\alpha$
cannot be equal to $1$ in four dimensional case and can, in no
way, be negative.

\section{Physical Features of the Present Model}
\subsection{Age of the Universe}
From equation (12) we find that if $\alpha=0$ (i.e. $\Lambda=0$),
then in four dimensional case for pressureless dust,
\begin{eqnarray}
t = \frac{2}{3H}.
\end{eqnarray}
It is clear from equation (16) that for the present accepted value
of the Hubble parameter, which is ($72\pm 8$)
kms$^{-1}$Mpc$^{-1}$~\cite{Altavilla2004}, the present age of the
Universe becomes less than the age of the globular clusters which
lies in the range $9.6$ Gyr~\cite{Bertolami1993}to $12.5\pm1.5$
Gyr~\cite{Gnedin2001,Cayrel2001}. This means that dark energy
models without $\Lambda$ suffer from low age problem. This
justifies the necessity for including the $\Lambda$ term in the
field equations as also mentioned by Sahni and
Starobinsky~\cite{Sahni2000}.

Again, for $n=2$ and $\alpha=1/3$, $Ht=1$. This implies that for
 pressureless dust, we can get the exact age of the present Universe which
 lies in the range ($14\pm 0.5$) Gyr~\cite{Spergel2003,Kirshner2003,Kunz2004,Tegmark2003}.
 It may be mentioned here that Ray and Mukhopadhyay~\cite{Ray2007b} obtained
 the exact age of the Universe for stiff fluid (i.e. $\omega=1$) for the same
 $\Lambda$ model with constant $G$. But stiff fluid model is not an accepted
 model for the present Universe. So, in that respect the present work has
 been successful in solving the so called age problem of the Universe
 without any unrealistic assumption. Moreover, it is easy to see that,
 with proper tuning of $\alpha$, the exact value of the present Universe
 can be obtained from equation (12) in dust case irrespective of the dimension.
 For instance, in five dimension (i.e. $n=3$), the accepted value of the age
 can be obtained for $\alpha=1$. Also for obtaining the exact age, the value of
 $\alpha=1$ has to be increased with the number of dimension and
 for each dimension the model suffers from low-age problem if we
 discard $\Lambda$, i.e. if we put $\alpha=0$.

\subsection{Rate of change of $G$}
From equation (14) we get,
\begin{eqnarray}
\frac{\dot G}{G} = -\frac{1}{t}.
\end{eqnarray}
Equation (17) tells us that the rate of change of $G$ is of the
order of $t^{-1}$. If we take the present age of the Universe as
$14$ Gyr, then (since $1$ Gyr is nearly $4\times 10^{17}$ seconds
and $1$ year is about $3\times 10^7$ seconds) the value of $\dot
G/G$ is of the order of $10^{-11}$ per year which is supported by
various theoretical and observational results (a comprehensive
list in this regard is provided in Ref. 22). Moreover, $\dot G/G$
does not depend on $n$, $\alpha$, mass of proton and Planck's
Constant. This means that the rate of change of $G$ is independent
of the space-time dimension, the parameter associated with dark
energy and two constants of microscopic world. This result
justifies the claim that dark energy does not exhibit any
gravitational effect like clustering and gravity is fundamentally
different from other types of forces of nature.

\subsection{Calculation of the deceleration parameter}
The deceleration parameter $q$ is given by
\begin{eqnarray}
q = -\frac{a \ddot a }{\dot a^2} = -\left(1+\frac{\dot
H}{H^2}\right).
\end{eqnarray}
Then, using equation (12), we get from equation (18),
\begin{eqnarray}
q = -\left[\frac{n(1+\omega)+(1+\omega)(6\alpha-n^2)}{2n}\right].
\end{eqnarray}
Then, for pressureless dust ($\omega=0$) we get from equation
(19),
\begin{eqnarray}
q = \left(\frac{n^2-n-6\alpha}{2n}\right).
\end{eqnarray}
It has been already shown that in four dimensional case ($n=2$) we
can get the exact age of the Universe, so far as observational
results are concerned, for $\alpha=1/3$. Now, for $n=2$ and
$\alpha=1/3$ we get $q=0$ which means that the Universe expands
with a uniform velocity. So, we are not getting an accelerating
Universe what the present cosmological scenario demands. Another
interesting point is, corresponding to every value of $\alpha$ for
which we get the exact age, we get the value of $q$ as zero
whatever may be the spatial dimension. This means that the
Universe expands with uniformly in every dimension. This situation
can be compared with the so called `hesitation period' (when the
quasars were supposed to be formed) of the
Eddington-Lema{\^i}tr{\'e} model. While in the
Eddington-Lema{\^i}tr{\'e} model the Universe behaves like
de-Sitter model for large $t$, here the Universe expands uniformly
for all the time. However, an accelerating Universe can be
obtained from this model also under proper tuning of~$\alpha$. For
example, in four dimensional case ($n=2$) with pressureless dust,
$q$ becomes $-0.1$ for $\alpha=0.4$ and in that case the age of
the Universe comes out as $14.9$ Gyr which is not an unreasonable
estimate. Similar is the case for all other dimensions.It may be
mentioned here that Khadekar and Butey~\cite{Khadekar2009} also
obtained $q=0$ for $N=1$ where under the assumption $\rho_m =
\gamma/R^N$, $\gamma$ and $N (>0)$ being constants It is
interesting to note here that for all those tuned values of
$\alpha$ which give us the exact age of the Universe, the scale
factor increases linearly. It may be mentioned here that linear
expansion in a Robertson-Walker Universe is shown to be possible
also by Usmani et al.~\cite{Usmani2008} for the $\Lambda$ model
$\dot \Lambda\sim H^3$.

\section{Conclusions}
In the present work, some new features of an widely used
$\Lambda$-dark energy model has been explored through an ($n+2$)
dimensional analysis with variable $\Lambda$ and variable $G$.
Recently Khadekar and Butey~\cite{Khadekar2009} have studied a
similar higher dimensional cosmological model where both $\Lambda$
and $G$ are variable. While Khadekar and Butey~\cite{Khadekar2009}
have chosen $G$ by following Eddington-Weinberg empirical
relation~\cite{Marugan2002} and Singh~\cite{Singh2006}, here in
the present case a particular relation of
Zeldovich's~\cite{Zeldovich1968} in connection to the cosmological
parameter is used for obtaining an expression for the so-called
gravitational constant. The approach of the present work is also
completely different from that of Khadekar and
Butey~\cite{Khadekar2009}.

In the present investigation we have been addressed, by using a
well known $\Lambda$ model, some key issues of modern cosmology,
viz. the age problem, the amount of variation of $G$ and the
nature of expansion of the Universe. The results we have shown are
as follows:

(i) For any dimension, the accepted age of the present Universe
can be attained under proper tuning of the parameter $\alpha$.

(ii) The amount of variation of the gravitational constant is
shown to be compatible with theoretically and observationally
established values without any prior assumption on the value of
the parameter $\alpha$. This is an improvement over the work of
Ray and Mukhopadhyay~\cite{Ray2007b} where the fine-tuning of the
parameters were essential for obtaining the currently accepted
values so far as the amount of variation of $G$ is concerned. But,
here no such precondition is required.

(iii) The present model has been successful in obtaining both
negative and zero values of the deceleration parameter $q$ which
correspond to accelerating and uniformly expanding Universe
respectively.

Finally, it should be mentioned that although in the present work
emphasis is given on the zero value of the equation of state
parameter $\omega$, but the present model also admits negative
values of $\omega$ with the restriction $0<\omega<-1$.

\section*{Acknowledgments}
One of the authors (SR) is thankful to the authority of
Inter-University Centre for Astronomy and Astrophysics, Pune,
India for providing Visiting Associateship under which a part of
this work was carried out.

\end{document}